\documentclass[aps,prb,twocolumn,showpacs,amsmath]{revtex4-1}
\usepackage{graphicx}
\usepackage{amsmath}
\usepackage{amsthm}
\begin{document}

\title{Why does bulk boundary correspondence fail in some non-hermitian
topological models}

\author{Ye Xiong}
\email{xiongye@njnu.edu.cn}
\affiliation{Department of Physics and Institute of Theoretical Physics
  , Nanjing Normal University, Nanjing 210023, P. R. China \\
   National Laboratory of Solid State Microstructures, Nanjing
  University, Nanjing 210093, P. R. China}

\begin{abstract}
  Bulk boundary correspondence is crucial  to topological insulator as
  it associates the boundary states (with zero energy, chiral or
  helical) to topological numbers defined in bulk. The application of
  this correspondence needs a prerequisite condition which is usually
  not mentioned explicitly: the boundaries themselves cannot alter the
  bulk states, so as to the topological numbers defined on them. In
  non-hermitian models with fractional winding number, we prove that such
  precondition fails and the bulk boundary correspondence is cut out. We
  show that, as eliminating the hopping between the boundaries to
  simulate the evolution of a system from the periodic boundary
  condition to the open boundary condition, exceptional points must be passed
  through and the topological structure of the spectrum has been
  changed. This makes the topological structures of a chain with open
  boundary totally different from that without the boundary. We also argue
  that such exotic behavior does not emerge when the open boundary is
  replaced by a domain-wall. So the index theorem can be applied to the
  systems with domain-walls but cannot be further used to those with
  open boundary.
\end{abstract}

\maketitle

\section{Indroduction}

In quantum mechanics of hermitian Hamiltonian, the degeneracy of the
energy spectrum plays crucial role in the generation of nontrivial topological
order, i.e., the nonzero (first kind of) Chern number is
generated by the effective magnetic monopoles at the degenerate points
in the parameter space\cite{Niu}. In a recent year, some authors try to spread these
ideas to the models with non-hermitian(NH) Hamiltonian\cite{Gong2010,
  Mailybaev2005, Rudner2009, Esaki2011, Zeuner2015, Lee2016,
Leykam2016b}. Besides the
topological phase that is smoothly extended from the hermitian
cases\cite{Esaki2011, PhysRevB.84.153101}, the NH models can possess
new topological phases stemming from a new kind of degenerate points,
the {\it exceptional points} (EPs)\cite{Moi, Berry2004, Heiss2012,
Mehri-Dehnavi2008, Liang2013, Malzard2015,  Cerjan2016, Lin2016a,
Xu2016e}.

The discussions on the NH Hamiltonian started more than half a century
ago and a special kind of NH models, $\mathcal{PT}$-symmetric models,
has been studied both theoretically and experimentally
\cite{RevModPhys.29.269, Bender1998, Bender2002, Mostafazadeh2002,
Cui2012,Lee2014, Malzard2015, Medvedyeva2016, Amir2015, Dattoli1990,
Mehri-Dehnavi2008, Muller2009a, Gania2010, Liang2013, Cui2014a,Lee2014b,
Bender2014a, Shah2015, Nixon2015, 0305-4470-38-9-L03,
PhysRevLett.100.103904, PhysRevLett.103.093902, PhysRevA.82.031801, PhysRevLett.103.123601,
Ruter2010, PhysRevA.82.043803, PhysRevLett.106.213901,
PhysRevA.85.050101, PhysRevLett.110.234101,PhysRevLett.113.023903,
Regensburger2013, Zhang2016g, Hodaei2014, PhysRevX.4.031042,
Fleury2015,Chen2016, Xiongd}.
EPs are the special points in a parameter space where the NH matrix
ceases to be diagonalizable because of the coalescences of the
eigenvalues and eigenstates. These properties can be
illustrated by the following $2\times2$ Jordan block reading as
\begin{equation}
H=\begin{pmatrix} 0 & 1 \\ r_0 e^{-ik} & 0 \end{pmatrix}.
  \label{eq1}
\end{equation}
The EP at $r_0=0$ is the point where the two eigenvalues coalesce to $0$ and the
right eigenstates (left eigenstates) coalesce to
$(1,0)^T$ ($(0,1)$). As there is only one eigenstate,
the Jordan block cannot be diagonalized any more at the EP. 
Each EP can induce a square root singularity so that there
are multiple square root branches in the parameter space around it.
This can also be illustrated with the above toy matrix by taking a positive
$r_0$ and encircling the EP by varying $k$ from $0$ to $2\pi$. The two
eigenvalues read 
\begin{equation}
  E_{\pm} = \sqrt{r e^{ik}}. 
\end{equation}
Due to the two branches in the complex plane induced by the square root,
the eigenvalues that continuously varying with $k$ will come back to
their original values after $4\pi$ period instead of the $2\pi$ period for the
matrix itself. This fact leads to the fractional winding number
introduced in the previous articles\cite{Lee2016, Leykam2016b}.

In the previous comment, we question that is it necessary to connect the
square root branches with the fractional winding number in a topological
language\cite{Xiongc}. In this article, we further prove that there is no
bulk-boundary correspondence in these NH models and the zero energy
boundary states (ZEBSs) are caused by the fact that the Hamiltonian is
right at (or exponentially close to) an EP when the boundary is open.
This is different from the ZEBSs that are protected by the chiral
symmetry in the traditional topological insulators. As the open boundary
is accompanying with EP while a domain-wall does not, the index theorem
presented in Ref. \onlinecite{Leykam2016b} can only be applied to the systems
with domain-wall but cannot be further extended to the systems with open
boundary condition (OBC). Besides that, many exotic properties emerge,
i.e., {\it all} bulk states are changed from extended states to the
exponentially localized states when the boundary condition is changed
from periodic to open and the bulk state spectrum are also entirely
changed during this process.

\section{Models and results}

We start this section with the toy matrix in Eq. \ref{eq1} because it
will illustrate many exotic features associated with EP. Some of the
methods used here can be applied to the general models.
We will talk about two kinds of EPs. When the Hamiltonian is
presented in the momentum space, the first kind of EPs is in the
parameter space spanned by $r_0$ and $k$ in Eq. \ref{eq1}, where $r_0$
is the hopping between the nearest neighboring unit cells and $k$ is the
wave-vector. While in the real space representation, the second kind of
EPs is in another parameter space spanned by $r$ and $\phi$ in Eq.
\ref{eqtr}, where $r$ is the hopping between the two ends of the chain
and $\phi$ is the phase added on this hopping. We hope that the readers will
not be confused by these two kinds.

\begin{figure}[htp]
  \centering
    \includegraphics[width=0.5\textwidth]{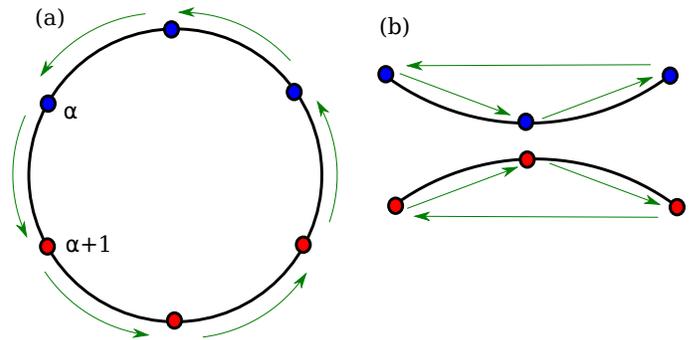}
    \caption{(a) The alternation of eigenvalues when $\phi$ is varying
      $2\pi$ continuously in Eq. \ref{eqtr}. As the translational
      symmetry is present in the case of $\phi=n2\pi$, the spectrum can be
      branched into $E_{\pm}$. Here the blue and red points
      represent the eigenvalues in these two branches, respectively.
      (b) For a typical hermitian Hamiltonian with two bands, inserting
      one quantum flux in the loop is equivalent to moving $k$ by
      $\frac{2\pi}{N}$ in the Brillouin zone. So the alternation of
    eigenvalues by varying $\phi$ occurs within each band and is different
from that in the NH case.}
  \label{fig1}
\end{figure}

The toy matrix can be considered as an effective NH Hamiltonian in the
momentum space for a 1-dimensional (1D) model. In the real space,
we suppose that the model is composited by $N$ unit cells
so that the Hamiltonian in the real space reads as
\begin{equation}
  H=\sum_{l=1}^N c^\dagger_{l,A}c_{l,B} +\sum_{l=1}^{N-1} r_0
  c^\dagger_{l,B} c_{l+1,A} + r e^{i\phi} c^\dagger_{N,B} c_{1,A},
  \label{eqtr}
\end{equation}
where $A$ and $B$ label the two inequivalent lattice sites in a unit
cell. Without loss of generality, we take $r_0$ to be real and positive.
The last term represents the hopping between the two ends. When $r=r_0$
and $\phi=0$, the translational symmetry restores and the spectrum can
be grouped into two branches by $E_{\pm} (k) = \pm \sqrt{r_0} e^{ik/2}$
with $N$ discrete $k$. In Fig. \ref{fig1} (a), we schematically show the
eigenvalues on a circle in the complex plane and the colors are used to
distinguish the two branches. For the sake of clarity in our next
discussion, we relabel these eigenvalues along the circle
counterclockwise by $E_{\alpha}$, where $\alpha=1,2,\cdots, 2N$. Then we
adjust $\phi$ from $0$ to $2\pi$ continuously in Eq. \ref{eqtr}. The
$2N$th-root of the complex number implies that $E_\alpha$ is
continuously changed to $E_{\alpha+1}$ and $E_{2N}$ is changed to $E_1$.
We want to emphasize that this pumping property is distinct from that in
the hermitian case and our general discussions will be based on it. In Fig.
\ref{fig1}, we schematically show this distinction.

\begin{figure}[htp]
  \centering
    \includegraphics[width=0.5\textwidth]{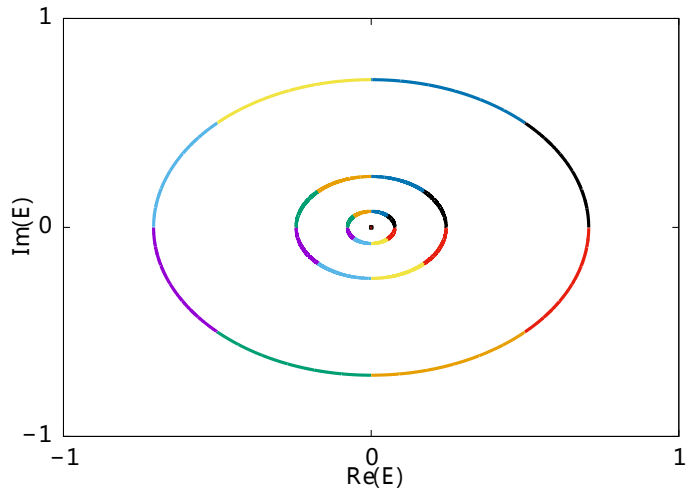}
    \caption{(a) From outside to inside, the traces in different colors
      show how the eigenvalues vary with $\phi$ in the complex
      plane when $r=r_0$, $10^{-4} r_0$, $10^{-8} r_0$ and $0$,
      respectively. The chain contains $N=4$ unit cells. So there are
      totally $8$ eigenvalues, which are represented by different
      colors in the figure. As long as $r\ne 0$, one needs $2N$ rounds
      of $\phi$ to return to the initial eigenvalue sheet. Here $r=0$ is
      an EP, where all eigenstates coalesce together.}
  \label{fig2}
\end{figure}

Next, we consider the effect of boundary by decreasing $r$ while varying
$\phi$ as usual.
In Fig. \ref{fig2}, we show how the eigenvalues evolve with $\phi$
in a $N=4$ chain when
$r=X^{2N}r_0$. In the figure, $X^{2N}$ is taken as $1$, $10^{-4}$, $10^{-8}$ and $0$, respectively.
Actually, When $r\ne r_0$, Eq. \ref{eqtr} can still be mapped back to a
translational symmetric matrix by a non-unitary transformation,
\begin{equation}
  H\to V^{-1} H V,
\end{equation}
with $V=\text{diag} (1,X,X^2,\cdots X^{2N-1})$. Here $V$ is a matrix
with only diagonal elements $1,X, \cdots, X^{2N-1}$. After the
transformation, the difference between the hopping amplitudes at the
boundary and in the bulk is smeared out and the hoppings in bulk are
rescaled by $X$.  This is why that only the radius but not the shape of
the circle is changed as $r$ is decreasing. We also notice that the
above transformation indicates that all the right eigenstates becomes
exponentially localized on the left end of the chain while the left
eigenstates localizes on the right end, which are confirmed by the
numerical calculations. 

Here $r=0$ is an EP of the Hamiltonian in Eq. \ref{eqtr}. Actually, this
is a $2N$ degenerate EP so that the $2N$ eigenvalues coalesce to $E=0$.
All the right and the left eigenstates coalesce to $(1,0,\cdots)^T$ and
$(0,\cdots,1)$, respectively. And we need to encircle this EP (by taking
$r\ne0$ and $\phi=0 \to 2\pi$) $2N$ rounds to reach the initial sheet of
the eigenstates. On should note that in the momentum space, the toy
model is encircling an EP as varying $k$ without touching any EP. But in
the real space representation, when the OBC is taken, the model is right
at an EP and the Hamiltonian matrix becomes defective. As the spectrum and
the eigenstates are changed entirely in approaching OBC, we have to
understand the ZEBSs from the coalescence of eigenstates at the EP instead of the
topological protection of boundary states caused by the bulk fractional winding
numbers. One reason is that the bulk spectrum has been dramatically changed
when the OBC is approached. This makes it impossible to connect the
topological band structure in the momentum space to the boundary states
in the real space because the two bulk spectra with and without OBC are totally different.
So the index theorems, such as the Thouless pump\cite{Ezaki2015}, cannot be applied any
more. We will present the other reasons after the studies of several
models.

In the momentum space, The 1D model in Ref. \onlinecite{Lee2016} is 
\begin{equation}
  H_k= (v+r_0\cos(k))\sigma_x + (r_0\sin(k)+ i \gamma/2)\sigma_z.
\end{equation}
After a unitary transformation $U=\frac{1}{\sqrt{2}}\begin{pmatrix} 1 & 1 \\ i &
-i \end{pmatrix}$, $H_k \to U^\dagger H_k U$, the Hamiltonian changes to 
\begin{equation}
  H_k= i \begin{pmatrix} 0 & (\gamma/2-v)-r_0e^{ik} \\
(\gamma/2+v)+r_0e^{-ik} & 0 \end{pmatrix}.
  \label{eq2}
\end{equation}
The EPs are at the points where either of the off-diagonal elements is
zero. We first take $\gamma=1$, $v=-0.5$ and $r_0=0.5$, which are in the
topological phase with fractional winding number in Ref.
\onlinecite{Lee2016}. Similar
to the toy model in the above discussion, 
we write down the Hamiltonian in the real space as
\begin{eqnarray}
  H & = & i\{\sum_{l=1}^N c^\dagger_{l,A}c_{l,B} +\sum_{l=1}^{N-1} 1/2
    [c^\dagger_{l,B} c_{l+1,A} - c^\dagger_{l+1,A} c_{l,B}] \nonumber \\
    & &  + r [e^{i\phi} c^\dagger_{N,B} c_{1,A}- e^{-i\phi} 
  c^\dagger_{1,A} c_{N,B}] \},
  \label{eq2r}
\end{eqnarray}
where $r$ and $\phi$ are the amplitude and the phase of the hopping between
the two ends of the chain, respectively. 

\begin{figure}[htp]
  \centering
    \includegraphics[width=0.5\textwidth]{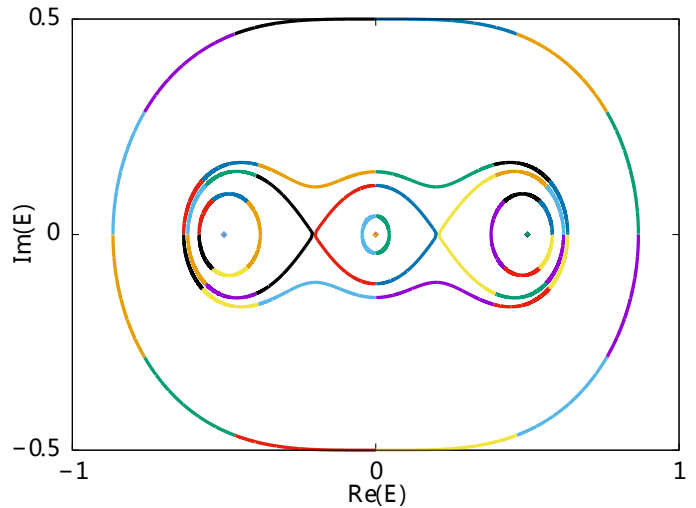}
    \caption{(a) The traces in different colors show how are the
      eigenvalues evolving when $\phi=0 \to 2\pi$. From outside to the
      inside, $r$ is taking $0.5$, $1\times 10^{-3}$, $5.23\times
      10^{-4}$, $6\times 10^{-5}$ and $0$, respectively. When
      $r=r_c=5.23\times 10^{-4}$, the Hamiltonian encounters an EP at
      which two pairs of eigenvalues coalesce. As further decreasing
      $r$, the alternation of eigenvalues splits into three unconnected
      loops. When $r=0$, there is another EP at where the eigenvalues
      coalesce to three points, $0$ and $\pm 0.5$. Here the length of
      the chain is $N=6$. Enlarging the chain does not change the
      evolution qualitatively, but $r_c$ will exponentially decrease to
  $0$.} 
  \label{fig3}
\end{figure}

In Fig. \ref{fig3}, we show how the energy spectrum is varying with
$\phi$ in a $N=6$ chain for several values of $r$. When $r$ is
relatively large, it
still needs totally $2N$ rounds to restore the initial sheet of the
eigenvalues because each round shifts the adjacent energy levels one by one
counterclockwise. When $r$ is smaller than a critical value, $r<r_c'$,
the circular alternation splits into three parts, in which two side
ones
include $N-1$ states and the center one has two states. So one will
need $2(N-1)$ rounds to reach the initial spectrum sheet when $N$ is an
even number or $(N-1)$ rounds when $N$ is odd. When $r$ is further
decreased to $0$, which corresponds to the chain with OBC,
the three circles shrink to three points at $\pm 0.5$
and $0$, respectively. So $r=0$ is also an EP of the Hamiltonian in the
real space. But the degeneracy of this EP is smaller than that in the
above toy model. Actually, there are totally three EPs that are
overlapping with each other at $r=0$,
whose degeneracies are $N-1$, $2$ and $N-1$, respectively. 
One should remember that there is also another EP at $(r=r_c',\phi=\pi)$, where two pairs of
eigenvalues coalesce. 

We also calculate the evolution of the spectrum for longer chains.
The circular alternations are similar to those in the short chain
presented in the above  figure but the EP at $r=r_c'$ is exponentially
rapidly moved to the EP at the origin as the length of the chain is
increased. 

\begin{figure}[htp]
  \centering
    \includegraphics[width=0.5\textwidth]{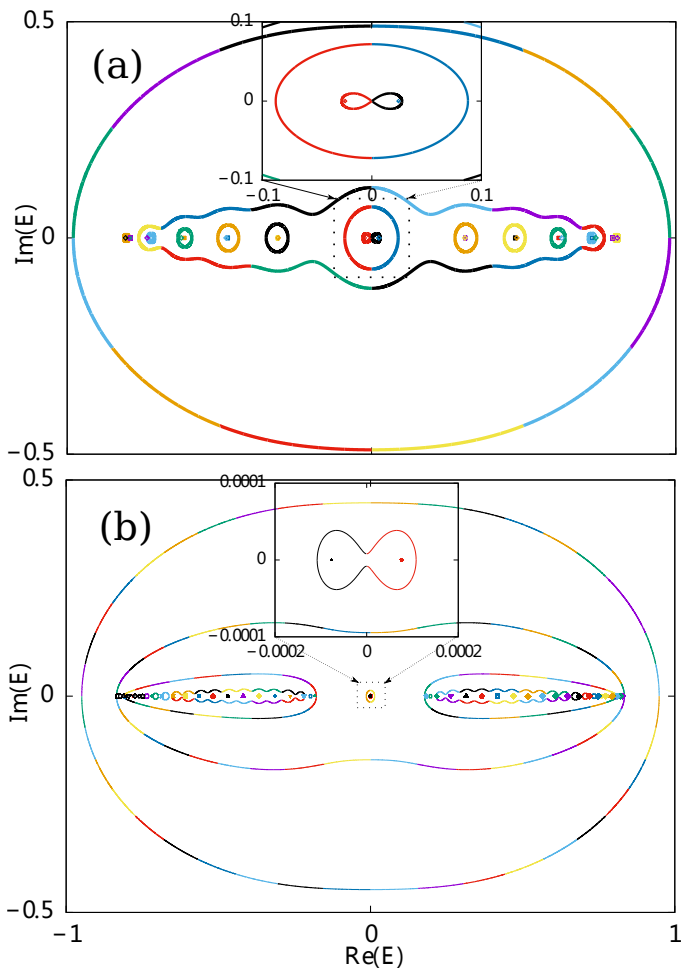}
    \caption{(a) The traces of the eigenvalues when the parameters are
  changed to $\gamma=1$, $v=-0.6$ and $r_0=0.5$ in Eq. \ref{eq2r}. Here
  from outside to the inside, $r$ is taking $0.5$, $1\times 10^{-3}$,
  $5\times 10^{-4}$, $3.2\times 10^{-5}$ and $0$, respectively. A zoom
  of the original region is shown in the inset. The length of the chain
  is still $N=6$. (b) The traces of the eigenvalues when the length of
  the chain is changed to $N=20$. Here $r$ is taking $0.5$, $1\times
  10^{-6}$, $1\times 10^{-9}$, $1\times 10^{-10}$, $5.3 \times 10^{-15}$
  and $0$, respectively. The inset shows the mini circle and the
  saturate points when $r=5.3 \times 10^{-15}$ and $0$.} 
  \label{fig4}
\end{figure}

We plot the results when the parameters are changed to $\gamma=1$,
$v=-0.6$ and $r_0=0.5$ in Fig. \ref{fig4}(a). In this case, the center EPs
at $r=0$ is split into many EPs and are moved away from the
origin. As Fig. \ref{fig4}(a) shows, when $r$ is decreased to $0.001$, one
EP has been encountered and the traces of eigenvalues are split into
three parts with a center large loop containing $10$ eigenvalues and
the two satellite circles each containing $1$ eigenvalue. As further
decreasing $r$, more EPs are encounted and more and more eigenvalues are
segregated from the center circle. When $r= r_c=3\times10^{-5}$, the last two eigenvalues at
the center coalesce. As all eigenvalues evolve to themselves when $r <
r_c$, there is no EP anymore. So in this model, $r=0$ is not an EP
and there are only two bound states near the zero energy when the OBC is
finally reached. But when the length of the chain is increased to
$N=20$, whose results have been shown in Fig. \ref{fig4}(b), all
EPs shrink toward $0$ rapidly. The last two adherent eigenvalues
coalesce at a much smaller $r_c \sim 5.3 \times 10^{-15}$ 
and stay more close to the zero energy
when $r=0$. So we can conclude that even $r=0$ is not an EP in this case,
all EPs are moving toward it exponentially rapidly with enlarging the system.

The 2-dimensional(2D) model in Ref. \onlinecite{Leykam2016b} reads as
\begin{equation}
  H_k= B_x(k_x,k_y)\sigma_z + B_y(k_x,k_y) \sigma_x,
\end{equation}
where $k_x$ and $k_y$ are the wave-vectors in the $x$ and $y$
directions, respectively. Although this is a 2D model, the authors
had considered $k_y$ as a parameter and characterize the topological
property by the fractional winding number in the $k_x$ direction. So
this model will be equivalent to the previous model in Eq.  \ref{eq2}
from the topological side of the view. 

\section{Discussions}

In real space representation, when the translational symmetry is
restored by taking the amplitude of the hopping between the ends equal to that
in bulk, $r=r_0$, the variation of $\phi$ by $2\pi$ is equivalent to
the shift of the wave-vector $k$ by $\frac{2\pi}{N}$ in the Brillouin zone. As
the period of the energy spectrum with $k$ is $4\pi$ in the momentum
space, their period with $\phi$ in the real space must be $4N\pi$
instead of $2N\pi$. So on the complex plane $(r\cos(\phi),r\sin(\phi))$,
there must have several EPs inside the circle $|re^{i\phi}|=r_0$. For
Eq. \ref{eq2r} with $\gamma=1$, $v=-0.5$ and $r_0=0.5$, after writing
down the Hamiltonian matrix with OBC, $r=0$, one can immediately realize
that this is an EP and there are at least two eigenvalues coalesce to
zero energy. We suggest to attribute this ZEBS to the EP instead to the
topological protected boundary state correspondent to the bulk
fractional winding numbers for the following reasons.

Firstly, as we mentioned previously, the spectrum of the
models with OBC or with periodic boundary condition are sharply
different. All the states, including the ZEBS, 
are exponentially localized at the boundary in the former
case, but are extended in the latter case. This distinction makes
the two systems uncorrelated so that the topological numbers defined in
the latter system has nothing to do with the spectrum when the OBC is
taken. 

Secondly, the fractional winding number defined in the momentum space is
stemmed from the $4\pi$ period of $k$. In real space, it has been
inherited by the $4N\pi$ period of $\phi$ when $r=r_0$. So we can
conclude that the topological number (fractional winding number here) is
encoded in the topology of the traces of the eigenvalues. To reach OBC
as decreasing $r$, one must encounter EPs and the topology of the traces
must be changed (as one large loop splits into smaller loops shown in
the previous figures).  So it is
impossible to associate the ZEBS at the open boundary to the fractional
winding number defined without boundary because the topologies of the
two systems are entirely different.

Thirdly, the ZEBSs are not protected by the chiral symmetry.  They are
actually caused by the fact that the Hamiltonian matrix with OBC has two
eigenvalues coalesce to zero energy or near the zero energy.  Taking Eq.
\ref{eq2} with the parameters $\gamma=1$, $v=-0.5$ and $r_0=0.5$ as an
example, we can destroy this EP at $r=0$ by adding the term
$h(c^\dagger_{1A} c_{1A} - c^\dagger_{NB}c_{NB})$, where $h$ is a
nonzero parameter. One should note that
this term only alter the on-site energies in the two boundary unit cells
but not the Hamiltonian in bulk. So if the ZEBSs are protected by the
winding number defined in the bulk of the chain, they will mostly not be
altered by the above extra term. But our numerical calculation indicates
that the above term destroys the ZEBSs. We can further support our
conclusion from another route.  The definition of the winding number in the
momentum space requires a chiral symmetry, which is $\sigma_z H_k
\sigma_z =-H_k$ in this article. If the topological understanding of
ZEBS is right, they must disappear when a term $ h\sigma_z$ is added in
the Hamiltonian in Eq. \ref{eq2}. In the real space representation, we
recover the EP at $r=0$ by eliminating the term $h(c^\dagger_{1A} c_{1A}
- c^\dagger_{NB}c_{NB})$ in the two unit cells at the boundaries.  A
simple numerical calculation confirms that the ZEBSs are still present.
So even when there is no chiral symmetry and the fractional winding
number is undefined, as long as the EP at $r=0$ is still present, the
ZEBSs can still exist. 

This article questions the topological understanding of ZEBS in NH
models. But we are not challenging most of the results in Ref.
\onlinecite{Leykam2016b} because the authors was discussing domain-walls instead of
free boundaries there. Unlike the models with open boundaries, a system
with domain-walls will not encounter the EP problem. But we want to
emphasize that their conclusions on the domain-walls can not be
further extended to the free boundaries. For instance, the index theorem
in that article starts from a translation $H'=H^\dagger H$ that maps the
NH Hamiltonian $H$ to an hermitian Hamiltonian $H'$. When $H$ is not
defective, the above translation maps the spectrum $\epsilon$ to
$|\epsilon|^2$ one by one. But the Hamiltonian $H$ will be defective
right at the EP so that the spectrum of $H'$ are not mapped one by one to
that of $H$ any more. The toy model in Eq. \ref{eq1} can illustrate
this: the eigenvalues of $H^\dagger H$ are not fixed at zero when
$r_0=0$. So the index theorem cannot be applied to the chain with OBC. 

\section{Conclusions}

We indicate that, as eliminating the amplitude of hopping between the
ends of a chain to reach OBC, EP must be passed through and the
topological structure of the band has been changed. This makes it
impossible to associate the ZEBS in the OBC case to the fractional
winding number defined without taking into account the boundary effect.
The topological index theorem on a domain-wall cannot be naturally
extended to that on the boundary for the same reason. The spectrum of
the chain with OBC should be studied individually and the topological
bulk boundary correspondence is cut out. Our studies also show that
there are EP at or exponentially adjacent to $r=0$ in a long chain in these
models. This makes it possible to study the effect of EP on a long chain
without finely tune the parameters. 

{\it Technical note}: Near or at the EP, the LU decomposition used by
the lapack subroutines, i.e., zgeev is not stable. So it will give a
wrong spectrum when the length of the chain is larger than $100$
typically. We use a bi-orthogonal Gram-Schmidt process to calculate the
spectrum in that case.


\end{document}